\documentclass[12pt,a4paper]{article}%
\usepackage{epsfig}
\usepackage{amsmath}
\usepackage{amsfonts}
\usepackage{amssymb}
\usepackage{graphicx}
\usepackage[T1]{fontenc}
\usepackage[utf8]{inputenc}
\usepackage{authblk}%
\providecommand{\U}[1]{\protect\rule{.1in}{.1in}}
\begin{document}

\title{Quantum Evolution of the Time-Dependent Non-Hermitian\ Hamiltonians: Real
Phases }
\author{{\small Mustapha Maamache}$^{a}$\thanks{E-mail: maamache@univ-setif.dz },
{\small Oum Kaltoum Djeghiour}$^{a,b}$\thanks{E-mail:
k.djeghiourjijel@gmail.com}, {\small Naima Mana}$^{a}$\thanks{E-mail:
na3ima\_mn@hotmail.fr} {\small and} {\small Walid Koussa}$^{a}$\thanks{E-mail:
koussawalid@yahoo.com}\\$^{(a)}${\small Laboratoire de Physique Quantique et Syst\`{e}mes Dynamiques,}\\{\small Facult\'{e} des Sciences, Universit\'{e} Ferhat Abbas S\'{e}tif 1,
S\'{e}tif 19000, Algeria}\textit{.}\\$^{(b)}${\small D\'{e}partement de Physique, Universit\'{e} de Jijel, } \\{\small BP 98 \ Ouled Aissa, 18000 Jijel, Algeria.} }
\date{}
\maketitle

\begin{abstract}
Explicitly time-dependent pseudo-Hermitian (TDPH) invariants theory systems,
with a time-dependent (TD) metric, is developed for a time-dependent non
Hermitian (TDNH) quantum systems. We derive a simple relation between the
eigenstates of this pseudo-Hermitien (PH) invariant and the solutions of the
Schrodinger equation. A physical system is treated in detail: \ the TD Swanson
model, where an explicitly TDPH invariant is derived for this system, the
eigenvalues and eigenstates of the invariant are calculated explicitly.

PACS: 03.65.-w, 03.65.Ca, 03.65.Ge

\end{abstract}

\section{Introduction}

Quantum mechanics is based on a set of axioms among them we mention: (i) the
inner products of state vectors have a positive norm, (ii) the time evolution
is unitary, iii) the Hamiltonian of a system must be a Hermitian operator
$h^{\dag}=h$ in order to guarantee that its eigenvalues are real. Recently,
Bender et al \cite{Carl1,Carl5} found that non-Hermitian (NH) Hamiltonians
satisfy these conditions and have interpreted the reality of the spectrum as
being due to its $\mathcal{PT}$-symmetry which comes from the invariance of
$\mathcal{PT}$-symmetric Hamiltonians under both parity and time reversal
transformation $H\ \mathcal{PT}=\mathcal{PT}H$ \ . The parity operator
$\mathcal{P}$ and the time reversal operator $\mathcal{T}$ are defined by
their action on the position operator $x$ and the momentum operator $p$ as:
$\mathcal{P}:x\rightarrow-x,\quad p\rightarrow-p,$ $\mathcal{T}$:
$x\rightarrow x,$ $p\rightarrow-p,$ $i\rightarrow-i\,$.

The crucial point is to redefine the inner product that enables to
re-establish the consistent probabilistic interpretation of the theory. Given
that in $\mathcal{PT}$ quantum mechanics, $\mathcal{P}$ and $\mathcal{T}$ take
a role analogous to the Hermitian conjugate in ordinary quantum mechanics, a
natural way to define the $\mathcal{PT}$ inner product of two eigenfunctions
of $H$ is given by $\langle\phi_{m}^{H}\left\vert \phi_{n}^{H}\right\rangle
_{\mathcal{PT}}=\left(  \mathcal{PT}\left\vert \phi_{m}^{H}\right\rangle
\right)  .\left\vert \phi_{n}^{H}\right\rangle =\left(  -1\right)  ^{n}%
\delta_{mn}$ which shows that this $\mathcal{PT}$ inner product is not always
definite positive. In quantum theory the inner product in Hilbert space of
state vectors has a positive norm. Positive definiteness is restored by
introducing a linear operator $\mathcal{C}$ that takes eigenstates of the
Hamiltonian that have negative norm under the $PT$ inner product and turns
them into positive \cite{Carl5}. Obviously, the operator $\mathcal{C}$ does
not commutes with $\mathcal{P}$ and $\mathcal{T}$ separately, but commutes
with the $\mathcal{PT}$ $\ $product. The $\mathcal{CPT}$ inner product which
is defined as
\[
\langle\phi_{m}^{H}\left\vert \phi_{n}^{H}\right\rangle _{CPT}=\left(
\mathcal{CPT}\left\vert \phi_{m}^{H}\right\rangle \right)  .\left\vert
\phi_{n}^{H}\right\rangle =\delta_{mn},
\]
has a positive norm.

It was established,\ in Refs. \cite{most1,most3,most4}, that $\mathcal{PT}$
symmetry is neither necessary nor sufficient for the reality of the spectrum
which can be attributed to the pseudo Hermiticity of the Hamiltonian $H$. A
Hamiltonian is called quasi-Hermitian \cite{Scholz} or pseudo-Hermitian (PH)
if it exists a bounded with respect to an invertible Hermitian operator
$\eta=\rho^{+}\rho$ satisfies%

\begin{equation}
H^{\dag}=\eta H\eta^{-1}. \label{1}%
\end{equation}
The NH Hamiltonian $H$ can be related to an equivalent Hermitian one by
\begin{equation}
h=\rho H\rho^{-1},
\end{equation}
showing that the eigenvalues of $h$ and $H$ are identical, although the
relations between their eigenvectors will differ
\begin{equation}
\left\vert \psi_{n}^{h}\right\rangle =\rho\left\vert \phi_{n}^{H}\right\rangle
.
\end{equation}
This, in turn, requires a redefinition of the usual inner product to
\begin{equation}
\langle\phi_{m}^{H}\left\vert \phi_{n}^{H}\right\rangle _{\eta}=\langle
\phi_{m}^{H}|\eta\left\vert \phi_{n}^{H}\right\rangle =\delta_{mn}.
\end{equation}

All these efforts have been devoted to study TDNH systems. Systems with TDNH
Hamiltonians operators and time-independent metrics have been studied in
\cite{Faria1,Faria2} , a number of conceptual difficulties have been
encountered in the generalization to TD metric operators
\cite{znojil1,most5,most6,znojil2,znojil3,Bila,wang1,wang2,mus,fring1,fring2,khant,frith,luiz1,luiz2}%
. Recent contributions \cite{fring1,fring2} have advanced that it is
incompatible to maintain unitary time evolution for TDNH Hamiltonians when the
metric operator is explicitly TD. In other words, both Hamiltonians $H(t)$ and
$H^{\dag}(t)$ involved are related to each other as%
\begin{equation}
H^{\dag}\left(  t\right)  =\eta\left(  t\right)  H\left(  t\right)  \eta
^{-1}\left(  t\right)  +i\hbar\dot{\eta}\left(  t\right)  \eta^{-1}\left(
t\right)  , \label{PHH1}%
\end{equation}
The key feature in this equation is the fact that the Hamiltonian $H(t)$ is no
longer quasi-Hermitian due to the presence of the last term and thus it
generalizes the well known standard quasi-Hermiticity relation $\left(
\ref{1}\right)  $ in the context TDNH quantum mechanics \cite{fring1,fring2}.

This work investigate in detail the main frames of TDNH systems ruled by the
Schr\"{o}dinger equation
\begin{equation}
H(t)\left\vert \Phi^{H}(t)\right\rangle =i\hbar\partial_{t}\left\vert \Phi
^{H}(t)\right\rangle , \label{shrod2}%
\end{equation}
where $\left\vert \Phi^{H}(t)\right\rangle $ is related to the solution
$\left\vert \Psi^{h}(t)\right\rangle $ of \ the Hermitian Schr\"{o}dinger
equation
\begin{equation}
h\left(  t\right)  \left\vert \Psi^{h}(t)\right\rangle =i\hbar\partial
_{t}\left\vert \Psi^{h}(t)\right\rangle , \label{PSCH}%
\end{equation}
by a TD invertible operator $\rho\left(  t\right)  $ as
\begin{equation}
\left\vert \Psi^{h}(t)\right\rangle =\rho\left(  t\right)  \left\vert \Phi
^{H}(t)\right\rangle . \label{vect}%
\end{equation}
For this we introduce, in section 2, the pseudo invariant operator theory for
the TD Schr\"{o}dinger equation related with the NH Hamiltonian. Then we give
the solution of theTD Schr\"{o}dinger equation in terms of eigenstates of PH
invariant operator $I^{PH}\left(  t\right)  $ and goes on to examine how the
reality of their phases can be established. \ In section 3, by using the
Lewis-Riesenfeld method \cite{Lewis} of invariants and a TD metric, we
construct a TD solutions for the generalized version of the NH Swanson
Hamiltonian with TD coefficients. Section 4, concludes our work.

\section{Pseudo-invariant operator method}

The use of invariants theory to solve quantum systems, whose Hamiltonian is an
explicit function of time, has the advantage to offer an exact solution for
problems solved by the traditional TD perturbation theory \cite{landau}. There
is a class of exact invariants for TD harmonic oscillators, both classical and
quantum, that has been reported in \cite{Lewis2}.

The invariants method \cite{Lewis} is very simple due to the relationship
between the eigenstates of the invariant operator and the solutions to the
Schr\"{o}dinger equation by means of the phases; in this case the problem is
reduced to find the explicit form of the invariant operator and the phases.

Now we proceed to introduce and analyze the spectral properties of PH
invariant operator $I^{PH}\left(  t\right)  $. Particular attention is given
to the special subset of quasi Hermitian operators. We start by considering a
NH quantum mechanics in its most general form by studying TD Hamiltonian
operators $H(t)$ satisfiying the Schr\"{o}dinger equation (\ref{shrod2}) and
where the metric operator $\eta\left(  t\right)  =\rho^{+}\left(  t\right)
\rho\left(  t\right)  $ associated with $H(t)$ and $I^{PH}\left(  t\right)  $
is also TD.

Suppose the existence of a pseudo Hermitian, explicitly TD, non trivial
invariant operator $I^{PH}\left(  t\right)  $;\ that means, $I^{PH}\left(
t\right)  $ satisfies%
\begin{equation}
I^{PH\dag}\left(  t\right)  =\eta(t)I^{PH}\left(  t\right)  \eta^{-1}(t)\text{
}\Leftrightarrow I^{h}(t)=\rho(t)I^{PH}(t)\rho^{-1}(t)=I^{h\dag}(t),
\label{quas}%
\end{equation}%
\begin{equation}
\frac{\partial I^{PH}(t)}{\partial t}=\frac{i}{\hbar}\left[  I^{PH}\left(
t\right)  ,H(t)\right]  . \label{LewisPH}%
\end{equation}
Thus $I^{PH}\left(  t\right)  $ may be mapped to $I^{h}\left(  t\right)  $, by
a similarity transformation $\rho(t)$.

It is easy to see that the action of the invariant operator in a
Schr\"{o}dinger state vector is also solution to the Schr\"{o}dinger equation,
that is%
\[
H(t)\left(  I^{PH}\left(  t\right)  \left\vert \Phi^{H}(t)\right\rangle
\right)  =i\hbar\partial_{t}\left(  I^{PH}\left(  t\right)  \left\vert
\Phi^{H}(t)\right\rangle \right)  ,
\]
which is a valid result for any invariant operator.

Now, we generalize the Lewis-Riesenfeld theory so that it can be used to find
the eigenstates $\left\vert \phi_{n}^{H}(t)\right\rangle $ of $I^{PH}(t)$%
\begin{equation}
\text{ \ }I^{PH}\left(  t\right)  \left\vert \phi_{n}^{H}(t)\right\rangle
=\lambda_{n}\left\vert \phi_{n}^{H}(t)\right\rangle , \label{IPH}%
\end{equation}
and%
\begin{equation}
\left\langle \phi_{m}^{H}(t)\right\vert \eta(t)\left\vert \phi_{n}%
^{H}(t)\right\rangle =\delta_{m,n}.
\end{equation}
The eigenvalues $\lambda_{n}$ are also time-independent, as we can deduce in
the following simple way. By differentiating Eq. (\ref{IPH}) with respect to
time, it follows that%
\begin{equation}
\text{ }\frac{\partial I^{PH}}{\partial t}\text{\ }\left\vert \phi_{n}%
^{H}(t)\right\rangle +\text{\ }I^{PH}\text{ }\frac{\partial\left\vert \phi
_{n}^{H}(t)\right\rangle }{\partial t}\text{\ }=\frac{\partial\lambda_{n}%
}{\partial t}\left\vert \phi_{n}^{H}(t)\right\rangle +\lambda_{n}%
\frac{\partial\left\vert \phi_{n}^{H}(t)\right\rangle }{\partial t},
\label{dif}%
\end{equation}
taking the scalar product of \ Eq. (\ref{dif}) with a state \ $\left\langle
\phi_{n}^{H}(t)\right\vert \eta(t)$ and using the left-hand side of Eq.
(\ref{LewisPH}), we obtain $\partial\lambda_{n}/\partial t=0$
\begin{equation}
\frac{\partial\lambda_{n}}{\partial t}=\left\langle \phi_{n}^{H}(t)\right\vert
\eta(t)\frac{\partial I^{PH}}{\partial t}\text{\ }\left\vert \phi_{n}%
^{H}(t)\right\rangle =0. \label{VP}%
\end{equation}
Since the Hermitian invariant $I^{h}(t)$ and the NH one $I^{PH}(t)$ are
related by a similarity transformation (\ref{quas}), therefore they have the
same eigenvalues.The reality of the eigenvalues $\lambda_{n}$ is guaranteed,
since one of the invariants involved, i.e. $I^{h}(t),$ is Hermitian.\ 

In order to investigate the connection between eigenstates of $I^{PH}\left(
t\right)  $ and solutions of the Schr\"{o}dinger equation,
\begin{equation}
i\hbar\partial_{t}\left\vert \Phi_{n}^{H}(t)\right\rangle =H(t)\left\vert
\Phi_{n}^{H}(t)\right\rangle ,
\end{equation}
we first start by projecting Eq. (\ref{dif}) onto $\left\langle \phi_{m}%
^{H}(t)\right\vert \eta(t)$ and using Eq. (\ref{VP}), we obtain%
\begin{equation}
i\hbar\left\langle \phi_{m}^{H}(t)\right\vert \eta(t)\frac{\partial}{\partial
t}\ \left\vert \phi_{n}^{H}(t)\right\rangle =\left\langle \phi_{m}%
^{H}(t)\right\vert \eta(t)H(t)\ \left\vert \phi_{n}^{H}(t)\right\rangle
,\ \ \ (m\neq n).
\end{equation}

The next step in the method is showing the existence of a simple and explicit
rule for choosing the phases of the eigenstates of $I^{PH}(t)$ such that these
states satisfy themselves the Schr\"{o}dinger equation with the only
requirement the invariant does not involve time differentiation. The new
eigenstates $\left\vert \Phi_{n}^{H}(t)\right\rangle $ of $I^{PH}(t)$ are%
\begin{equation}
\left\vert \Phi_{n}^{H}(t)\right\rangle =e^{i\gamma_{n}(t)}\left\vert \phi
_{n}^{H}(t)\right\rangle , \label{sol}%
\end{equation}
will satisfy the Schrodinger equation.This is to say, $\left\vert \Phi_{n}%
^{H}(t)\right\rangle $ is particular solution to the Schrodinger equation.
This requirement is equivalent to the following first-order differential
equation for the $\gamma_{n}(t)$:%
\begin{equation}
\frac{d\gamma_{n}(t)}{dt}=\left\langle \phi_{n}^{H}(t)\right\vert
\eta(t)\left[  i\hbar\frac{\partial}{\partial t}-H(t)\right]  \text{\ }%
\left\vert \phi_{n}^{H}(t)\right\rangle . \label{Phase}%
\end{equation}
In Eq. (\ref{Phase}), the first term is parallel to a familiar non-adiabatic
geometrical phase, but the second term representing effects due to a
time-dependent Hamiltonian is a dynamical phase. It is the sum of these two
terms that can ensure a real $\gamma_{n}(t)$.

The general solutions of the Schrodinger equation for system with
non-Hermitian time-dependent Hamiltonian $H(t)$ are readily obtained as
follows:%
\begin{equation}
\left\vert \Phi^{H}(t)\right\rangle =%
{\textstyle\sum_{n}}
C_{n}e^{i\gamma_{n}(t)}\left\vert \phi_{n}^{H}(t)\right\rangle ,
\label{Gensol}%
\end{equation}
where the $C_{n}$ = $\left\langle \phi_{n}^{H}(0)\right\vert \eta(0)\left\vert
\Phi^{H}(0)\right\rangle $ are time-independent coefficients.

\section{Generalized time dependent non-Hermitian Swanson Hamiltonian}

The first model of a NH $\mathcal{PT}$-symmetric Hamiltonian quadratic in
position and momentum was studied by Ahmed \cite{ahmed} and made popular by
Swanson \cite{swanson} namely $H=\omega\left(  a^{+}a+\frac{1}{2}\right)
+\alpha a^{2}+\beta a^{+2}$ with $\omega,\alpha$ and $\beta$ real parameters,
such that $\alpha\neq$ $\beta$ and \ $\omega^{2}-4\alpha\beta>0$ and where
$a^{+}$ and $a$ are the usual harmonic oscillator creation and annihilation
operators for unit frequency. This Hamiltonian has been studied extensively in
the literature by several authors \cite{jones,bagchi,musumbu,quesne,sinha,eva}.

We construct here, by employing the Lewis-Riesenfeld method of invariants, the
solutions for the non-Hermitian Swanson Hamiltonian with TD coefficients
\cite{fring2}
\begin{equation}
H(t)=\omega(t)\left(  a^{+}a+\frac{1}{2}\right)  +\alpha(t)a^{2}%
+\beta(t)a^{+2}, \label{HH}%
\end{equation}
\ where $\left(  \omega(t),\alpha(t),\beta(t)\right)  $ $\in C$ are
time-dependent parameters. We set $\hbar=1$.

Following the previous idea, the problem is reduced to find a pseudo invariant
operator. The most general invariant $I^{PH}(t)$, for the generalized Swanson
oscillator (\ref{HH}), can be written in the form
\begin{equation}
I^{PH}(t)=\delta_{1}(t)\left(  a^{+}a+\frac{1}{2}\right)  +\delta_{2}%
(t)a^{2}+\delta_{3}(t)a^{+2}, \label{IH}%
\end{equation}
where $\delta_{1}(t)$, $\delta_{2}(t)$, $\delta_{3}(t)$ are time dependent
real parameters. The invariant (\ref{IH}) is of course manifestly NH when
$\delta_{2}(t)\neq$ $\delta_{3}(t)$.

Let us solve the standard quasi-Hermiticity relation (\ref{quas}) by making
the following general and, for simplicity, Hermitian ansatz for a TD metric
$\rho\left(  t\right)  $%
\begin{align}
\rho\left(  t\right)   &  =\exp\left[  \epsilon\left(  t\right)  \left(
a^{+}a+\frac{1}{2}\right)  +\mu\left(  t\right)  a^{2}+\mu^{\ast}\left(
t\right)  a^{+2}\right]  ,\nonumber\\
&  =\exp\left[  \vartheta_{+}\left(  t\right)  K_{+}\right]  \exp\left[
\ln\vartheta_{0}\left(  t\right)  K_{0}\right]  \exp\left[  \vartheta
_{-}\left(  t\right)  K_{-}\right]  , \label{metr}%
\end{align}
where $K_{+}=a^{+2}/2,$ $K_{-}=a^{2}/2,$ $K_{0}=\left(  a^{+}a/2+1/4\right)  $
form SU(1, 1)-algebra
\begin{equation}
\left\{
\begin{array}
[c]{c}%
\left[  K_{0},K_{+}\right]  =K_{+}\\
\left[  K_{0},K_{-}\right]  =-K_{-}\\
\left[  K_{+},K_{-}\right]  =-2K_{0}%
\end{array}
\right.  , \label{su(1,1)}%
\end{equation}
with the TD coefficients
\begin{align}
\vartheta_{+}\left(  t\right)   &  =\frac{2\mu^{\ast}\sinh\theta}{\theta
\cosh\theta-\epsilon\sinh\theta}=-\Phi(t)e^{-i\varphi(t)},\nonumber\\
\vartheta_{0}\left(  t\right)   &  =\left(  \cosh\theta-\frac{\epsilon}%
{\theta}\sinh\theta\right)  ^{-2}=\Phi^{2}(t)-\chi(t),\label{TDC}\\
\vartheta_{-}\left(  t\right)   &  =\frac{2\mu\sinh\theta}{\theta\cosh
\theta-\epsilon\sinh\theta}=-\Phi(t)e^{i\varphi(t)},\nonumber\\
\chi(t)  &  =-\frac{\cosh\theta+\frac{\epsilon}{\theta}\sinh\theta}%
{\cosh\theta-\frac{\epsilon}{\theta}\sinh\theta}\text{ \ \ \ \ \ \ ,\ \ \ }%
\theta=\sqrt{\epsilon^{2}-4\left\vert \mu\right\vert ^{2}}.\nonumber
\end{align}
\ The key point is the construction of the Hermitian invariant operator
$I^{h}(t)=\rho(t)I^{PH}(t)\rho^{-1}(t)$ from the NH one $I^{PH}(t)$. It
follows that
\begin{align}
I^{h}(t)  &  =\frac{2}{\vartheta_{0}}\left[  \left[  -\delta_{1}\left(
\vartheta_{-}\vartheta_{+}+\chi\right)  -2\left(  \delta_{2}\vartheta
_{+}+\delta_{3}\chi\vartheta_{-}\right)  \right]  K_{0}\right. \nonumber\\
&  \left.  +\left(  \delta_{1}\vartheta_{-}+\delta_{2}+\delta_{3}\vartheta
_{-}^{2}\right)  K_{-}+\left(  \delta_{1}\chi\vartheta_{+}+\delta_{2}%
\vartheta_{+}^{2}+\delta_{3}\chi^{2}\right)  K_{+}\right]  . \label{invh}%
\end{align}
The equation (\ref{invh}) has been derived with the help of the following
relations%
\begin{align}
\rho\left(  t\right)  K_{+}\rho^{-1}\left(  t\right)   &  =\frac{1}%
{\vartheta_{0}}\left[  -2\vartheta_{-}\chi K_{0}+\vartheta_{-}^{2}K_{-}%
+\chi^{2}K_{+}\right]  ,\nonumber\\
\rho\left(  t\right)  K_{0}\rho^{-1}\left(  t\right)   &  =\frac{1}%
{\vartheta_{0}}\left[  -\left(  \vartheta_{-}\vartheta_{+}+\chi\right)
K_{0}+\vartheta_{-}K_{-}+\chi\vartheta_{+}K_{+}\right]  ,\\
\rho\left(  t\right)  K_{-}\rho^{-1}\left(  t\right)   &  =\frac{1}%
{\vartheta_{0}}\left[  -2\vartheta_{+}K_{0}+K_{-}+\vartheta_{+}^{2}%
K_{+}\right]  .\nonumber
\end{align}

For $I^{h}(t)$ to be Hermitian ($I^{h}(t)=I^{\dag h}(t)$) we require the
coefficient of $K_{0}$ is real, and the coefficients of $K_{-}$ and $K_{+}$
are complex conjugate of one another. Using these two requirements, we have:%
\begin{align}
\left[  -\delta_{1}\left(  \vartheta_{-}\vartheta_{+}+\chi\right)  -2\left(
\delta_{2}\vartheta_{+}+\delta_{3}\chi\vartheta_{-}\right)  \right]   &
=\left[  -\delta_{1}\left(  \vartheta_{-}\vartheta_{+}+\chi\right)  -2\left(
\delta_{2}\vartheta_{-}+\delta_{3}\chi\vartheta_{+}\right)  \right]
,\nonumber\\
\left(  \delta_{1}\vartheta_{-}+\delta_{2}+\delta_{3}\vartheta_{-}^{2}\right)
&  =\left(  \delta_{1}\chi\vartheta_{-}+\delta_{2}\vartheta_{-}^{2}+\delta
_{3}\chi^{2}\right)  ,\\
\left(  \delta_{1}\chi\vartheta_{+}+\delta_{2}\vartheta_{+}^{2}+\delta_{3}%
\chi^{2}\right)   &  =\left(  \delta_{1}\vartheta_{+}+\delta_{2}+\delta
_{3}\vartheta_{+}^{2}\right)  ,\nonumber
\end{align}
which correspond to
\begin{align}
\delta_{2}  &  =\delta_{3}\chi,\nonumber\\
\delta_{1}  &  =-\frac{\delta_{3}\left(  \vartheta_{-}^{2}+\chi\right)
}{\vartheta_{-}}=-\frac{\delta_{3}\left(  \vartheta_{+}^{2}+\chi\right)
}{\vartheta_{+}}. \label{delta}%
\end{align}
From equation (\ref{delta}), it follows that $\vartheta_{+}=$ $\vartheta
_{-}\equiv-\Phi(t)$ implying that the TD parameter $\mu(t)$ must be real, i.e.
$\mu(t)=\mu^{\ast}(t)$. Finally the similarity transformation\ (\ref{metr})
maps the NH quadratic invariant \ (\ref{IH}) into $I^{h}(t)$ given by%
\begin{equation}
I^{h}\left(  t\right)  =-\frac{2}{\vartheta_{0}}\left[  \delta_{1}\left(
\Phi^{2}+\chi\right)  -4\delta_{3}\chi\Phi\right]  K_{0}. \label{invh1}%
\end{equation}

Let $\left\vert \psi_{n}^{h}\right\rangle $ be the eigenstate of $K_{0}$ with
the eigenvalue $k_{n}$ i.e.%
\begin{equation}
K_{0}\left\vert \psi_{n}^{h}\right\rangle =k_{n}\left\vert \psi_{n}%
^{h}\right\rangle .
\end{equation}
The eigenstates of $I^{h}\left(  t\right)  $ (\ref{invh1}) are obviously given
by
\begin{equation}
I^{h}\left(  t\right)  \left\vert \psi_{n}^{h}(t)\right\rangle =-\frac
{2}{\vartheta_{0}}\left[  \delta_{1}\left(  \Phi^{2}+\chi\right)  -4\delta
_{3}\chi\Phi\right]  k_{n}\left\vert \psi_{n}^{h}\right\rangle \text{, \ }%
\end{equation}
because of the time-dependence, the invariant $I^{h}\left(  t\right)  $ is a
conserved quantity whose eigenvalues are real constants. However, without loss
of generality, the factor $-\left[  \delta_{1}\left(  \Phi^{2}+\chi\right)
-4\delta_{3}\chi\Phi\right]  /\vartheta_{0}$ can be taken equal to $1$.\ It
follows that the eigenstates $\left\vert \phi_{n}^{H}(t)\right\rangle $ of
$I^{PH}(t)$ can be directly deduced from the basis $\left\vert \psi_{n}%
^{h}\right\rangle $ of its Hermitian counterpart $I^{h}\left(  t\right)  $
through the similarity transformation $\left\vert \phi_{n}^{H}(t)\right\rangle
=$ $\rho^{-1}(t)\left\vert \psi_{n}^{h}\right\rangle $ with time-independent
eigenvalue $k_{n}$.

According to the above discussion, the problem is reduced to find a PH
invariant operator and the suitable real phases of its eigenfunctions to take
them as the solution for the Schr\"{o}dinger equation. In a first step, we
will determine the real parameters $\delta_{1},\delta_{2},\delta_{3}$ so that
our invariant operator $I^{PH}(t)$ (\ref{IH}) is PH. Imposing the quasi-
Hermiticity condition (\ref{quas}) on $I^{h}\left(  t\right)  $, we get%
\begin{align}
I^{\dag PH}(t)  &  =\rho^{+}\left(  t\right)  I^{h}\left(  t\right)
\rho^{-1+}\left(  t\right)  =2\delta_{1}K_{0}+2\delta_{3}K_{-}+2\delta
_{2}K_{+}\nonumber\\
&  =-\frac{2}{\vartheta_{0}}\left[  \left(  \Phi^{2}+\chi\right)  K_{0}+\Phi
K_{-}+\chi\Phi K_{+}\right]  .
\end{align}
From the above equation, the real parameters $\delta_{1},\delta_{2},\delta
_{3}$ follow straightforwardly:%
\begin{equation}
\delta_{1}=-\frac{\left(  \Phi^{2}+\chi\right)  }{\vartheta_{0}}\text{ ,
}\delta_{2}=-\frac{\chi\Phi}{\vartheta_{0}}\text{ , }\delta_{3}=-\frac{\Phi
}{\vartheta_{0}}.
\end{equation}
Therefore, the PH invariant operator $I^{PH}(t)$ is written in the following
form%
\begin{equation}
I^{PH}(t)=-\frac{2}{\vartheta_{0}}\left[  \left(  \Phi^{2}+\chi\right)
K_{0}+\chi\Phi K_{-}+\Phi K_{+}\right]  . \label{PH1}%
\end{equation}

The second step in the method is imposing for $I^{PH}(t)$(\ref{PH1}) the
invariance condition (\ref{LewisPH}) which leads to the following relations:%
\begin{equation}
\dot{\vartheta}_{0}=\frac{\vartheta_{0}}{\Phi}\left[  -2\Phi\left\vert
\omega\right\vert \sin\varphi_{\omega}+\left\vert \alpha\right\vert
\sin\varphi_{\alpha}+\left(  2\Phi^{2}+\chi\right)  \left\vert \beta
\right\vert \sin\varphi_{\beta}\right]  , \label{cont1}%
\end{equation}%
\begin{equation}
\dot{\Phi}=-\Phi\left\vert \omega\right\vert \sin\varphi_{\omega}+\left\vert
\alpha\right\vert \sin\varphi_{\alpha}+\Phi^{2}\left\vert \beta\right\vert
\sin\varphi_{\beta}, \label{cont2}%
\end{equation}%
\begin{equation}
\
\begin{array}
[c]{c}%
\chi\left\vert \beta\right\vert \cos\varphi_{\beta}=\left\vert \alpha
\right\vert \cos\varphi_{\alpha}\\
\left(  \Phi^{2}+\chi\right)  \left\vert \alpha\right\vert \cos\varphi
_{\alpha}=\chi\Phi\left\vert \omega\right\vert \cos\varphi_{\omega}\\
\Phi\left\vert \omega\right\vert \cos\varphi_{\omega}=\left(  \Phi^{2}%
+\chi\right)  \left\vert \beta\right\vert \cos\varphi_{\beta}%
\end{array}
, \label{rel}%
\end{equation}
here, $\varphi_{\omega}$, $\varphi_{\alpha}$ and $\varphi_{\beta}$ are the
polar angles of $\omega$, $\alpha$, and $\beta$, respectively.

The final step consists in determining the Schrodinger solution (\ref{sol})
which is an eigenstate of the PH invariant (\ref{PH1}) multiplied by a
time-dependent factor (\ref{Phase})%
\begin{align}
\frac{d\gamma_{n}(t)}{dt}  &  =\left\langle \phi_{n}^{H}(t)\right\vert
\eta(t)\left[  i\frac{\partial}{\partial t}-H(t)\right]  \text{\ }\left\vert
\phi_{n}^{H}(t)\right\rangle \nonumber\\
&  =\left\langle \psi_{n}^{h}\right\vert \left[  i\rho\dot{\rho}^{-1}-\rho
H\rho^{-1}\right]  \text{\ }\left\vert \psi_{n}^{h}\right\rangle .
\label{Phase1}%
\end{align}
Using the NH Hamiltonian $H(t)$ (\ref{HH}) and then deriving the transformed
Hamiltonian $\left[  i\rho\dot{\rho}^{-1}-\rho H\rho^{-1}\right]  $\ through
the metric operator $\rho(t)$ (\ref{metr}), we further identify this
transformed Hamiltonian as
\begin{equation}
i\rho\dot{\rho}^{-1}-\rho H\rho^{-1}=2W\left(  t\right)  K_{0}+2U\left(
t\right)  K_{-}+2V\left(  t\right)  K_{+},
\end{equation}
where the coefficient functions are%
\begin{align}
W\left(  t\right)   &  =\frac{1}{\vartheta_{0}}\left[  \omega\left(  \Phi
^{2}+\chi\right)  -2\Phi\left(  \alpha+\beta\chi\right)  -\frac{i}{2}\left(
\dot{\vartheta}_{0}-2\Phi\dot{\Phi}\right)  \right]  ,\\
U\left(  t\right)   &  =\frac{1}{\vartheta_{0}}\left[  \omega\Phi-\alpha
-\beta\Phi^{2}+\frac{i}{2}\dot{\Phi}\right]  ,\\
V\left(  t\right)   &  =\frac{1}{\vartheta_{0}}\left[  \omega\chi\Phi
-\alpha\Phi^{2}-\beta\chi^{2}+\frac{i}{2}\left(  \vartheta_{0}\dot{\Phi}%
+\Phi^{2}\dot{\Phi}-\Phi\dot{\vartheta}_{0}\right)  \right]  .
\end{align}
Considering Eqs. (\ref{rel}), these TD coefficients can be simplified as%
\begin{align}
W\left(  t\right)   &  =\frac{1}{\vartheta_{0}}\left[  \left\vert
\omega\right\vert \left(  \Phi^{2}+\chi\right)  \cos\varphi_{\omega}%
-4\Phi\left\vert \alpha\right\vert \cos\varphi_{\alpha}\right. \nonumber\\
&  \left.  -i\frac{\vartheta_{0}}{2\Phi}\left[  -\Phi\left\vert \omega
\right\vert \sin\varphi_{\omega}+\left\vert \alpha\right\vert \sin
\varphi_{\alpha}+\chi\left\vert \beta\right\vert \sin\varphi_{\beta}\right]
\right]  ,\\
U\left(  t\right)   &  =0,\\
V\left(  t\right)   &  =0.
\end{align}
Knowing that the phase $\gamma_{n}(t)$ (\ref{Phase1}) must be real, we need to
impose that the frequency $W\left(  t\right)  $ is real. Then, we obtain the
exact phase of the eigenstate%
\begin{equation}
\gamma_{n}(t)=k_{n}%
{\displaystyle\int\limits_{0}^{t}}
\frac{2}{\vartheta_{0}}\left[  \left\vert \omega\right\vert \left(  \Phi
^{2}+\chi\right)  \cos\varphi_{\omega}-4\Phi\left\vert \alpha\right\vert
\cos\varphi_{\alpha}\right]  dt^{\prime}.
\end{equation}
Therefore, the solutions for the Schr\"{o}dinger equation (\ref{shrod2}) are
given by
\begin{equation}
\left\vert \Phi^{H}(t)\right\rangle =%
{\textstyle\sum_{n}}
C_{n}(0)\exp\left(  ik_{n}%
{\displaystyle\int\limits_{0}^{t}}
\frac{2}{\vartheta_{0}}\left[  \left\vert \omega\right\vert \left(  \Phi
^{2}+\chi\right)  \cos\varphi_{\omega}-4\Phi\left\vert \alpha\right\vert
\cos\varphi_{\alpha}\right]  dt^{\prime}\right)  \left\vert \phi_{n}%
^{H}(t)\right\rangle .
\end{equation}
The canonical representation (\ref{PH1}), when is expressed in terms of
$\ x=\frac{1}{\sqrt{2}}(a+a^{+})$ and $p=\frac{i}{\sqrt{2}}(a^{+}-a)$, becomes%
\begin{equation}
I^{PH}(t)=\frac{1}{2\vartheta_{0}}\Big\{\left[  \left(  \Phi-\chi\right)
\left(  1-\Phi\right)  \right]  p^{2}-i\Phi\left(  \chi-1\right)  \left(
px+xp\right)  -\left[  \left(  \Phi+\chi\right)  \left(  1+\Phi\right)
\right]  x^{2}\Big\},
\end{equation}
and the eigenfunctions are given as%

\begin{align}
\phi_{n}^{H}(x,t)  &  =\sqrt{\frac{1}{n!2^{n}\sqrt{\pi}}\sqrt{\frac
{\vartheta_{0}}{\left(  \Phi-\chi\right)  \left(  1-\Phi\right)  }}}%
\exp\left[  -\frac{1}{2}\left(  \frac{\vartheta_{0}+\Phi\left(  \chi-1\right)
}{\left(  \Phi-\chi\right)  \left(  1-\Phi\right)  }\right)  x^{2}\right]
\nonumber\\
&  \times H_{n}\left(  \left[  \frac{\vartheta_{0}}{\left(  \Phi-\chi\right)
\left(  1-\Phi\right)  }\right]  ^{\frac{1}{2}}x\right)  .
\end{align}
Clearly the eigenvalues are $2k_{n}=\left(  n+1/2\right)  $ and the $H_{n}$
are the Hermite polynomials of order $n$. These eigenfunctions are orthonormal
with respect to the weight \ factor $\eta\left(  t\right)  =\exp\left[
\frac{\Phi\left(  \chi-1\right)  }{\left(  \Phi-\chi\right)  \left(
1-\Phi\right)  }x^{2}\right]  $. That is,%

\begin{equation}
\int\phi_{m}^{\ast H}(x,t)\exp\left[  \frac{\Phi\left(  \chi-1\right)
}{\left(  \Phi-\chi\right)  \left(  1-\Phi\right)  }x^{2}\right]  \phi_{n}%
^{H}(x,t)dx=\delta_{mn}.
\end{equation}

\section{Conclusion}

Recently, the general framework for a description of a unitary time evolution
for TDNH Hamiltonians has been stated and the use of a TD metric operator
cannot ensure the unitarity of the time evolution simultaneously with the
observability of the Hamiltonian \cite{fring1,fring2}. They adapted the method
based on a TD unitary transformation of TD Hermitian Hamiltonians
\cite{mizrahi,mus1} to solve the Schr\"{o}dinger equation for the generalized
version of the NH Swanson Hamiltonian withTD coefficients.

In this work, using quasi-Hermiticity relation (\ref{quas}) between a NH
invariant operator $I^{PH}(t)$ and Hermitian one $I^{h}(t)$, we have presented
an alternative approach to solve the time-evolution of quantum systems. We
investigated in detail the main frames ofTD systems in the framwork of the
Lewis and Riesenfeld method which ensures that a solution to the
Schr\"{o}dinger equation governed by a TDNH Hamiltonian is an eigenstate of an
associated PH invariant operator $I^{PH}(t)$ with a TD global real phase
factor $\gamma_{n}(t)$.

The properties derived here help us to understand better systems described by
TDNH Hamiltonians and should play a central role in TDNH quantum mechanics.
After going through these properties, we then have presented an illustrative
example: the generalized Swanson Hamiltonian with TD complex coefficients.

\end{document}